\documentclass[12pt,letterpaper]{article}
\pdfoutput=1

\usepackage{amsmath,amssymb,calc, amsthm,bbm}
\usepackage{graphicx}
\usepackage{hyperref}
\usepackage[nosort]{cite}
\usepackage{epsfig,psfrag}
\usepackage{mathtools}

\makeatletter
\renewcommand\section{\@startsection {section}{1}{\z@}%
                                   {-3.5ex \@plus -1ex \@minus -.2ex}
                                   {2.3ex \@plus.2ex}%
                                   {\normalfont\large\bfseries}}
\renewcommand\subsection{\@startsection{subsection}{2}{\z@}%
                                     {-3.25ex\@plus -1ex \@minus -.2ex}%
                                     {1.5ex \@plus .2ex}%
                                     {\normalfont\bfseries}}
\makeatother

\theoremstyle{plain}

\theoremstyle{definition}

\let\non\nonumber

\def\one{^{(1)}}

\newcommand{\bea}{\begin{eqnarray}}
\newcommand{\eea}{\end{eqnarray}}
\newcommand{\be}{\begin{equation}}
\newcommand{\ee}{\end{equation}}
\newcommand{\bma}{\begin{pmatrix}}
\newcommand{\ema}{\end{pmatrix}}

\newcommand{\Z}{{\mathbb Z}}
\newcommand{\R}{{\mathbb R}}

\newcommand{\wt}{\widetilde}

\newcommand{\La}{\Lambda}

\newcommand{\G}{\Gamma}
\newcommand{\e}{\epsilon}

\newcommand{\com}[2]{{ \left[ #1, #2 \right] }}
\newcommand{\acom}[2]{{ \left\{ #1, #2 \right\} }}

\newcommand{\m}{\mu}
\newcommand{\n}{\nu}

\def\com#1#2{{ \left[ #1, #2 \right] }}
\def\acom#1#2{{ \left\{ #1, #2 \right\} }}

\newcommand{\C}[1]{$(\ref{#1})$}

\def\IZ{\relax\ifmmode\mathchoice
{\hbox{\cmss Z\kern-.4em Z}}{\hbox{\cmss Z\kern-.4em Z}}
{\lower.9pt\hbox{\cmsss Z\kern-.4em Z}} {\lower1.2pt\hbox{\cmsss
Z\kern-.4em Z}}\else{\cmss Z\kern-.4em Z}\fi}
\def\IR{\relax{\rm I\kern-.18em R}}

\def\one{{\hbox{ 1\kern-.8mm l}}}

\newlength{\bredde}
\def\slash#1{\settowidth{\bredde}{$#1$}\ifmmode\,\raisebox{.15ex}{/}
\hspace*{-\bredde} #1\else$\,\raisebox{.15ex}{/}\hspace*{-\bredde}
#1$\fi}

\newsavebox{\zzzbar}
\sbox{\zzzbar}
  {\setlength{\unitlength}{0.9em}
  \begin{picture}(0.6,0.7)
  \thinlines
  \put(0,0){\line(1,0){0.6}}
  \put(0,0.75){\line(1,0){0.575}}
  \multiput(0,0)(0.0125,0.025){30}{\rule{0.3pt}{0.3pt}}
  \multiput(0.2,0)(0.0125,0.025){30}{\rule{0.3pt}{0.3pt}}
  \put(0,0.75){\line(0,-1){0.15}}
  \put(0.015,0.75){\line(0,-1){0.1}}
  \put(0.03,0.75){\line(0,-1){0.075}}
  \put(0.045,0.75){\line(0,-1){0.05}}
  \put(0.05,0.75){\line(0,-1){0.025}}
  \put(0.6,0){\line(0,1){0.15}}
  \put(0.585,0){\line(0,1){0.1}}
  \put(0.57,0){\line(0,1){0.075}}
  \put(0.555,0){\line(0,1){0.05}}
  \put(0.55,0){\line(0,1){0.025}}
  \end{picture}}

\newcommand{\ena}{\end{eqnarray}}
\newcommand{\beqa}{\begin{eqnarray}}
\newcommand{\eeqa}{\end{eqnarray}}

\def\G{\Gamma}

\def\H{{\cal H}}

\def\bes #1\ees{\begin{split}#1\end{split}}

\newfont{\goth}{ygoth.tfm scaled 1200}                   

\def\e{\epsilon}
\def\th{\theta}

\def\m{\mu}
\def\n{\nu}

\def\G{\Gamma}

 \numberwithin{equation}{section}

\def\1{{(1)}}
\def\2{{(2)}}
\def\3{{(3)}}

\def\1{{\bf 1}}

\def\M{{\mathcal M}}

\def\1{{\bf 1}}
\def\3{{\bf 3}}
\def\7{{\bf 7}}
\def\2{{\bf 2}}
\def\8{{\bf 8}}

\newcommand{\lp}{\left(}
\newcommand{\rp}{\right)}

\newcommand{\om}{\omega}

\newcommand{\hph}[1]{{\hphantom{#1}}}

\renewcommand{\wt}{\widetilde}

\newcommand{\ov}{\overline}
\newcommand{\ul}{\underline}

\def\bes #1\ees{\begin{split}#1\end{split}}

\begin{document}
\begin{titlepage}

\begin{center}

{December 12, 2015}
\hfill         \phantom{xxx}  EFI-15-31, MI-TH-1544

\vskip 2 cm {\Large \bf A Landscape of Field Theories} 
\vskip 1.25 cm {\bf Travis Maxfield$^{a}$, Daniel Robbins$^{b}$ and Savdeep Sethi$^{a}$}\non\\
\vskip 0.2 cm
 $^{a}${\it Enrico Fermi Institute, University of Chicago, Chicago, IL 60637, USA}
\vskip 0.2 cm
$^{b}${\it George P. and Cynthia W. Mitchell Institute for Fundamental Physics and Astronomy, Texas A\&M University, College Station, TX 77843-4242, USA
}

\vskip 0.2 cm
{ Email:} \href{mailto:maxfield@uchicago.edu}{maxfield@uchicago.edu}, \href{mailto:drobbins@physics.tamu.edu}{drobbins@physics.tamu.edu}, \href{mailto:sethi@uchicago.edu}{sethi@uchicago.edu}

\end{center}
\vskip 1 cm

\begin{abstract}
\baselineskip=18pt

Studying a quantum field theory involves a choice of space-time manifold and a choice of background for any
 global symmetries of the theory. 
We argue that many more choices are possible when specifying the background.
In the context of branes in string theory, the additional data corresponds to a choice of supergravity tensor fluxes. We propose the existence of a landscape of field theory backgrounds, characterized by the space-time metric, global symmetry background and a choice of tensor fluxes. As evidence for this landscape, we study the supersymmetric  six-dimensional $(2,0)$ theory compactified to two dimensions. Different choices of metric and flux give rise to distinct two-dimensional theories, which can preserve differing amounts of supersymmetry.

\end{abstract}

\end{titlepage}

\tableofcontents

\section{Introduction} \label{intro}

This project is concerned with two closely related questions. The first question is quite classic: imagine a supersymmetric quantum field theory in $D$ space-time dimensions. Can one classify the interesting rigid backgrounds for this theory? Usually the term ``interesting'' implies backgrounds that preserve some degree of supersymmetry. We will see that string theory suggests a more general class of interesting backgrounds. The term ``background'' usually refers to a choice of space-time metric and a choice of background for any global symmetries of the field theory. Again we will see that string theory suggests a more general notion of background. 

The second question involves a special class of backgrounds, distinguished because they give rise to the degrees of freedom of a lower-dimensional quantum field theory in a Kaluza-Klein reduction. Included in this class are compact spaces, but the class is broader and can include non-compact spaces like a Taub-NUT manifold. This reduction provides a natural association between geometry and the lower-dimensional field theory 
derived by compactifying the parent theory. We will see that string theory suggests the existence of a landscape of field theories obtained from a parent theory by compactification on backgrounds that involve a choice of metric and a choice of flux.  

There has been much progress on both these questions in recent years, starting with supersymmetric sphere compactifications and the use of localization to compute partition functions~\cite{Pestun:2009nn}. That spheres can preserve supersymmetry is surprising, and counter to the usual intuition that a supersymmetric background requires a killing spinor.  The nicest explanation for the preserved supersymmetry comes from coupling the supersymmetric field theory to a non-dynamical background supergravity theory~\cite{Festuccia:2011ws}. From this perspective, backgrounds admitting killing spinors are on-shell supergravity configurations. Many examples of this type can be found in string theory and M-theory by studying branes wrapping submanifolds of a ten or eleven-dimensional metric. 

Backgrounds that do not admit killing spinors, but still preserve some supersymmetry, can be understood as off-shell supergravity configurations for which the supergravity auxiliary fields are not required to satisfy their equations of motion. We will argue that the beautiful structures that have been uncovered so far are part of a much larger landscape of supersymmetric and non-supersymmetric backgrounds. The generalization we have in mind is quite analogous to the discovery of flux vacua in string theory, which greatly enlarged the space of supersymmetric string vacua~\cite{Becker:1996gj, Dasgupta:1999ss}. 

Both closely related questions are framed entirely within quantum field theory. However, it is going to be useful for us to support our claim by studying a particular class of examples that follow from M-theory. We will consider an M5-brane wrapping a $4$-cycle of a Calabi-Yau $4$-fold. The M5-brane supports the $D=6$ $(2,0)$ theory. The theory has a $Spin(5,1)_L \times Spin(5)_R$ Lorentz and R-symmetry group. Kaluza-Klein reduction gives a string supporting a $D=2$ field theory. The preserved supersymmetry on the string, along with the spectrum of excitations, will depend on the following data: the topological type of the Calabi-Yau space, the choice of metric for the space, and the choice of supergravity fluxes. 

Since this example is derived from M-theory, it falls into the on-shell category. Despite this restriction, we will still see evidence for a landscape of $D=2$ theories obtained by including the flux degrees of freedom. We will comment on the relation between on-shell and off-shell backgrounds for theories with maximal supersymmetry at the close of this introduction. The reduction of the $(2,0)$ theory to $D=2$ has a long history, including applications to black hole entropy~\cite{Maldacena:1997de, Minasian:1999qn}, and more recently a prominent role in the program of associating geometry to field theory~\cite{Gadde:2013sca, Gadde:2013lxa, Benini:2013cda, Bak:2015taa}. 

The paper is organized as follows: in section~\ref{fluxbackgrounds}, we describe a class of M-theory flux vacua. In certain cases, there exist heterotic dual theories, which we also describe. This duality between certain M-theory flux backgrounds and heterotic string vacua provides the basic evidence for the field theory landscape. It will also help set our expectations for the kind of $D=2$ sigma model geometry that can emerge from wrapped M5-branes in the presence of flux. 

From the perspective of field theory, we understand how to introduce a background metric $g_{\m\n}.$ The metric couples to the field theory stress-energy tensor $T_{\m\n}$. We would like a similar understanding for background fluxes. Namely, which operators are activated in the field theory? Unlike the background metric case, the flux couples to operators which are tensors under both the $Spin(5,1)_L$ Lorentz symmetry and the $Spin(5)_R$ symmetry. It is natural to expect the operators to be part of the stress-tensor supermultiplet, constructed using the full $(2,0)$ superconformal symmetry. In section~\ref{hardwork}, we  answer the question of how the flux couples to the $(2,0)$ theory by studying an M5-brane coupled to $11$-dimensional supergravity. We describe the fixing of kappa symmetry and the emergence of a world-volume field theory, the preserved world-volume supersymmetries, and we examine some aspects of the resulting equations of motion.  Some of the early work on the coupling of fluxes to branes appears in~\cite{Kallosh:2005yu, Martucci:2005rb}\ and for Euclidean M5-branes in~\cite{Tsimpis:2007sx}.

There are a couple of observations and future directions worth mentioning in closing: 
\begin{itemize}
\item We study the on-shell case with fluxes that emerges from M-theory. However, we expect an off-shell extension to exist to some degree. Interestingly, it is unclear to what degree. By this we mean the following: theories with maximal supersymmetry currently do not admit fully off-shell superspace formalisms, with manifest maximal supersymmetry. This is true both for field theory and supergravity. This is unlike the case of half-maximal supersymmetry nicely studied in recent work~\cite{Butter:2015tra}. The closest to an appropriate off-shell supergravity formalism in $D=6$ is the $(2,0)$ conformal supergravity formulated in~\cite{Bergshoeff:1999db}. This formalism has been used to describe backgrounds for $D=5$ supersymmetric Yang-Mills~\cite{Cordova:2013cea}. 

Our initial comparison of on-shell fluxes with off-shell auxiliary fields suggests that a subset of configurations that might be called off-shell supergravity configurations appear to be related to on-shell flux backgrounds with auxiliary fields identified with fluxes. The extent to which this is true is unclear, but it is something that needs to be understood more precisely. 

\item   The computation of supersymmetric partition functions has been a significant advance in recent years~\cite{Pestun:2009nn}. We would like to be able to compute such partition functions on more general backgrounds that involve flux. There are hints that this might be possible. For example, it is not hard to construct many $D=2$ theories with $(0,2)$ or more world-sheet supersymmetry. For such models, the elliptic genus is a computable quantity. Whether it, or a higher-dimensional generalization, is computable from the $D=6$ $(2,0)$ theory perspective is an outstanding question. 

\item While we focus on reductions from $D=6$ to $D=2$, we expect the landscape enlargement to be much broader. There are already many more on-shell examples from string theory flux solutions to explore. It is not hard to imagine flux generalizations of compactifications with differing amounts of supersymmetry from $D=6$ to $D=4$ or lower dimensions, from $D=4$ to $D=2$, and so on. For example, potentially generalizing the $D=6$ to $D=4$ work of~\cite{Bah:2012dg}. 

\item Supersymmetric YM theories studied on rigid backgrounds give rise to interesting classes of vacuum equations; for example, Hitchin equations in various contexts. The solution spaces of the vacuum equations can be very interesting. A prime motivation for this work was to find vacuum equations for M5-branes in flux backgrounds because the $D=6$ theory provides a tensor generalization of the SYM vacuum structure. In M-theory examples that possess heterotic dual descriptions, we can build the heterotic string from the wrapped M5-brane and its target space geometry is expected to capture both heterotic torsion as well as heterotic non-geometric physics~\cite{Sethi:2007bw, McOrist:2010jw, Andriot:2011iw, Gu:2014ova, Malmendier:2014uka}. This would be an alternative to doubled field theory as an approach to defining non-geometric backgrounds~\cite{Hull:2009mi, Aldazabal:2013sca}. 

\item The classification of supersymmetric solutions of supergravity, including flux, is the study of G-structures~\cite{Gauntlett:2001ur}. What we are proposing is a field theory version of the supergravity classification. There are fewer constraints on the field theory side. For example, many of the key tadpole constraints found in string theory and M-theory are relaxed. We might therefore reasonably expect interesting new classes of backgrounds in field theory.  

\item Most of our discussion involves choices of flux that preserve supersymmetry. However, one can also consider a choice that solves the SUGRA equations of motion but does not preserve supersymmetry. If a truncation to supergravity is permissible, namely that higher momentum couplings can be ignored, then turning on this kind of flux gives a background which is special from the bulk perspective, since it breaks supersymmetry at a scale below the Kaluza-Klein scale. From a purely field theory perspective, this construction suggests we should enlarge our notion of ``interesting'' backgrounds to include these special classes of non-supersymmetric backgrounds.  
 
\end{itemize}

\vskip 0.5 cm \noindent {\bf Note Added:} During the final completion of this project, an interesting paper appeared which also discusses some aspects of branes in flux backgrounds~\cite{Triendl:2015lka}.

\section{Flux Backgrounds}\label{fluxbackgrounds}

\subsection{M-theory on a CY $4$-fold}

Our motivation for considering more general field theory backgrounds that involve fluxes comes from string theory.  We will begin by reviewing the structure of M-theory flux vacua based on Calabi-Yau $4$-folds~\cite{Becker:1996gj}. There are many other cases one might consider but we will focus on this case because it provides a great deal of intuition about how to include flux degrees of freedom. 

Let us label $10$ or $11$-dimensional bosonic coordinates by $M, N, \ldots$ and spinor indices by $a,b, \ldots$. We use the term ``bosonic coordinates" here because  we will later meet superspace coordinates which will denote by $A, B, \ldots$.  Let $\M_8$ denote a Calabi-Yau $4$-fold. The background of interest is a warped product of $\R^{2,1} \times \M_8$ with metric, 
\be
ds^2 = e^{2w(y)} \eta dx^2 + e^{-w(y)} \, ds^2_{\M_8}(y). 
\ee
Here $w(y)$ is the warp factor which depends on the internal coordinates collectively denoted by $y$. The Minkowski metric for $\R^{2,1}$ is denoted by $\eta dx^2$.  

In addition to the manifold, we choose an internal space $4$-form flux $G_4^{\rm int}$. At a basic level, the cohomology class of the flux satisfies the quantization condition~\cite{Witten:1996md}
\be\label{quantize}
\left[ {G_4^{\rm int}\over 2\pi}\right] - {p_1(\M_8)\over 4} \in H^4(\M_8,\Z). 
\ee
However, we stress that supergravity does not see this quantization condition. At the level of supergravity, we can smoothly turn on and off fluxes.  As we turn off any flux, the warp factor $w(y)$ tends to zero and $\M_8$ becomes Ricci flat. The last ingredient we need is a $3$-form potential, where $G_4=dC_3$ defines the potential. The purely space-time components of the potential are given by, \be C_3 = f(y)\, {\rm vol}_3,\ee where ${\rm vol}_3$ is the volume form of $\R^{2,1}$. We will fix $f(y)$ momentarily via equations of motion. 

Supersymmetry, along with the supergravity equation of motion,
\be\label{fluxeom}
d\star G_4 = -{1\over 2} G_4 \wedge  G_4,
\ee
requires that 
\be\label{self-duality} G_4^{\rm int} = {\hat\star}_8 G_4^{\rm int}, \ee 
where hatted quantities refer to the unwarped Calabi-Yau metric and $\hat\star_8$ is the Hodge dual restricted to the Calabi-Yau directions. In addition,  equation~\C{fluxeom}\ relates $f(y)$ and the warp factor
\be
f(y) =  - e^{3w(y)},
\ee  
and also fixes the warp factor in terms of the flux:
\be\label{warpfactor}
\hat\Delta(e^{-3w}) ={\hat\star}_8\left\{ -{1\over 2}  G^{\rm int}\wedge G^{\rm int} \right\}.
\ee
The Laplacian appearing in~\C{warpfactor}\ is again defined with respect to the unwarped Calabi-Yau metric. 

The right hand side of~\C{warpfactor}\ vanishes only if $G^{\rm int}=0$. If this is not the case, we have net membrane charge on $\M_8$, which requires $\M_8$ to be non-compact. This is the usual no-go result prohibiting flux on a compact space in $11$-dimensional supergravity. If we want to work strictly within the confines of supergravity then we should impose~\C{warpfactor}\ and consider only non-compact spaces $\M_8$. This is not a terrible restriction from the perspective of wrapped branes. The field theory on the brane really only sees the local normal bundle to the submanifold on which it is wrapped. It effectively only sees a local geometry, rather than the entire compact space. 

There are two ways to modify~\C{warpfactor}\ to go beyond the restriction imposed by supergravity. The first is to include $8$ derivative couplings in the M-theory effective action~\cite{Becker:1996gj}, denoted $X_8$ below, and the second is to include $N_{\rm M2}$ explicit M2-branes~\cite{Sethi:1996es}. This leads to the modified warp factor equation, 
\be
\hat\Delta(e^{-3w}) ={\hat\star}_8\left\{ -{1\over 2}  G^{\rm int}\wedge G^{\rm int} + 4\pi^2 X_8 - 4\pi^2\sum_{i=1}^n \delta^8(y-y_i) \right\}.
\ee  
For compact $\M_8$, the integrability condition for this equation requires vanishing of the total membrane charge:
\be
{1\over 2} \int_{\M_8} {G^{\rm int}\over 2\pi} \wedge {G^{\rm int} \over 2\pi} + N_{\rm M2} = {1\over 24} \chi(\M_8). 
\ee

The Calabi-Yau background preserves $4$ real supersymmetries, which we call $\mathcal{N}=2$ in 3 dimensions. The self-duality condition~\C{self-duality}\ and quantization condition~\C{quantize}\ requires that a generic $G^{\rm int}$ is, typically, an integral class with Hodge type: 
$$(4,0)\oplus (2,2)\oplus (0,4).$$ 
A generic flux breaks all the background supersymmetries in a manner that can be viewed as spontaneous breaking for sufficiently large Calabi-Yau volume. Here we again need to differentiate the case of compact $\M_8$ from the case of a non-compact $\M_8$. In the compact case, neither supersymmetric fluxes nor non-supersymmetric fluxes give solutions to the supergravity equations of motion. One must take into account higher derivative corrections. 

However, for supersymmetric backgrounds, there is good reason to expect a solution to the M-theory equations of motion in a perturbative  expansion in the $11$-dimensional Planck length $\ell_p$. In the non-supersymmetric case, there are equally good reasons to expect that a solution does not exist. This point will be elaborated on elsewhere. For non-supersymmetric fluxes, we really should restrict to non-compact spaces, where a truncation to two derivative supergravity is self-consistent. 

Preserving the full $\mathcal{N}=2$ supersymmetry requires a $(2,2)$ flux, which is also primitive with respect to the K\"ahler form $J$ of the Calabi-Yau~\cite{Dasgupta:1999ss}:
\be
J\wedge G^{\rm int}=0. 
\ee 
One can also preserve $\mathcal{N}=1$ supersymmetry by a choice of flux that is neither primitive nor $(2,2)$ but rather of the special form~\cite{Berg:2002es, Prins:2013wza}
\be
G^{\rm int} = c_1 \left( J\wedge J + {3\over 2} {\rm Re}(\Omega) \right) + G^{(2,2)}, 
\ee
with $c_1$ a constant and $\Omega$ the holomorphic top form of $\M_8$. This flux can also be viewed as spontaneously breaking $\mathcal{N}=2$ to $\mathcal{N}=1$. 

Aside from these generic possibilities, we could also consider a space $\M_8$ with a holonomy group properly contained in $SU(4)$ so the space itself preserves more  than $\mathcal{N}=2$ supersymmetry. The amount of supersymmetry preserved by the background would then depend on the particular choice of flux~\cite{Dasgupta:1999ss}. A nice case of this type is $\M_8=K3\times K3$, which will play a role in our subsequent discussion. This example preserves ${\mathcal N}=4$. To summarize this brief review: $11$-dimensional supergravity admits a large set of non-compact Calabi-Yau flux vacua which preserve either $\mathcal{N}=0,1$ or $2$ supersymmetries in $3$ dimensions. Further, there is evidence for a large set of compact Calabi-Yau flux vacua in M-theory, which preserve some degree of supersymmetry. 

\subsection{Heterotic flux vacua}

There are many generalizations of the flux story but for our purpose of understanding wrapped M5-branes, we only need to consider one other class of flux vacua. Consider a compactification of the heterotic or type I string to $3$ dimensions with flux.  For simplicity, we will use the language of the heterotic string. 
The general form of $3$-dimensional heterotic flux vacua has not yet been described, though there has been considerable progress in~\cite{Gauntlett:2002sc, Gauntlett:2003cy}. Fortunately, we can get by with an understanding of certain special cases that include dual descriptions of M-theory on $K3$-fibered spaces $\M_8$. 

Finding compact examples of Minkowski or $AdS_3$ heterotic torsional backgrounds is a difficult task. The known Minkowski examples correspond to spaces, $\M_7$, which take the form of a $3$-torus fibration over a base $\M_4$ with metrics:
\be\label{torsionalmetric}
ds^2_H =\eta dx^2 + \sum_{i=1}^3 e^{u_i(y)} \left( d\th_i + A_i \right)^2 +  \, ds^2_{\M_4}(y). 
\ee
The connections $A_i$ determine the structure of the $3$-torus fibration, where the torus itself has coordinates $(\th_1, \th_2, \th_3)$. There is also, typically, a dilaton that varies with $y$ as well as $H_3$-flux. We are being fairly schematic about the form of these metrics because the detailed structure is not important for us. 

What is important for us is a duality relating M-theory on a space $\M_8$ that admits a $K3$-fibration to the heterotic string on a torsional space with a metric of the form~\C{torsionalmetric}~\cite{Dasgupta:1999ss, Becker:2009df}. This duality follows from the more basic $7$-dimensional duality relating M-theory on a $K3$ surface to the heterotic string on $T^3$ to which we turn momentarily. 

First there is a special case that deserves mention. Some spaces $\M_7$ take the form $\M_7 = \M_6\times S^1$ with the $S^1$ isometry extending to a symmetry of both the $H_3$-flux and the dilaton. Essentially, one of the circle fibrations in~\C{torsionalmetric}\ is trivial. In these cases, we can decompactify the circle to find a  $4$-dimensional compactification of the heterotic string. If minimal $\mathcal{N}=1$ space-time supersymmetry is preserved then the two-dimensional world-sheet theory has $(0,2)$ supersymmetry.  
If an M-theory dual exists for this case, it corresponds to a geometry, $\M_8$, which admits an elliptic fibration, along with a choice of flux compatible with taking an F-theory limit. 

\subsection{Constructing the dual heterotic string}\label{constructhet}

Let us revisit the duality between M-theory on a $K3$ surface and the heterotic string on $T^3$. The generic space-time gauge fields visible on the heterotic side arise in M-theory from reducing the $3$-form potential, $C_3$, of $11$-dimensional supergravity on the lattice of integral $2$-forms of the $K3$ surface. This Kaluza-Klein reduction gives rise to $22$ gauge-fields associated to the cohomology lattice $\G^{3,3}\oplus\G^{16,0}.$ From the heterotic perspective, the lattice $\G^{3,3}$ is associated to $T^3$ while $\G^{16,0}$ is associated to the ten-dimensional gauge-fields.

To construct the dual heterotic string, we wrap an M5-brane on the $K3$ surface. The reduction of the M5-brane action on a $K3$ surface has been studied quite extensively~\cite{Cherkis:1997bx, Park:2009me}.  Later in our discussion, we will extend those analyses to the case of more general $4$-submanifolds of $\M_8$, including flux. 

For the moment, we want to gain some intuition about wrapped M5-branes and flux so a more qualitative discussion will suffice. Including only the lowest order interactions in  $\ell_p$, the bosonic light modes supported on an M5-brane consist of $5$ scalars $\phi^I$ and a $2$-form gauge-field $B_2$ with self-dual $3$-form field strength: 
\be
\H_3 = dB_2, \qquad  \H_3 = \ast \H_3.  
\ee 
There are also $16$ chiral fermions. Together these fields enjoy $(2,0)$ supersymmetry in $6$ dimensions. For a single M5-brane, the theory is free.  To see how the massless modes of the dual heterotic string emerge from the $(2,0)$ theory, reduce $B_2$ on the lattice of $2$-forms of the $K3$ surface. The result is $22$ chiral bosons. These bosons assemble themselves into $(16,0)$ chiral bosons of definite chirality, and $(3,3)$ chiral bosons corresponding to the $3$ non-chiral scalars parametrizing a $T^3$. 

The heterotic world-sheet can be presented in many ways. To study interacting world-sheets, it is usually convenient to describe the $(16,0)$ degrees of freedom in terms of left-moving world-sheet fermions coupled to the space-time gauge bundle,  along with $8$ non-chiral scalars and their right-moving fermion superpartners, which couple to the target space geometry and flux. Supersymmetry acts only on the right moving degrees of freedom with these conventions. The fundamental heterotic string only requires minimal $(0,1)$ supersymmetry in order to couple the theory to world-sheet supergravity. For the case of toroidal compactification, the world-sheet theory enjoys $(0,8)$ extended world-sheet supersymmetry. 

We want to make two observations about this well-known duality. The first concerns the coupling to space-time gauge fields. In analogy to the $F_2-B_2$ combination found on D-branes, the gauge-invariant combination of $B_2$ and $C_3$ found on M5-branes takes the form,
\be
\H_3 = dB_2 -C_3.  
\ee     
 After Kaluza-Klein reduction on the $i^{th}$ 2-cycle of $K3$, the resulting chiral boson $\phi^i$ couples to the corresponding space-time gauge-field via the combination $d\phi^i - A^i$. If we fermionize the $(16,0)$ chiral bosons, the result is a collection of chiral fermions coupled to a space-time gauge-bundle. The world-sheet theory is an interacting non-linear sigma model. Indeed if we do not include flux degrees of freedom, we would be unable to construct the most heavily studied class of heterotic world-sheets from the wrapped M5-brane! This is one direct motivation for considering flux backgrounds. 
 
 The second comment concerns parameters. M-theory on $K3$ depends on two basic scales: $\ell_p$ and the volume $V_4$ of the $K3$ surface. From these two parameters, one can determine the heterotic string length $\ell_s$ and the $7$-dimensional string coupling $g_s$; see~\cite{Becker:2007zj}\ for a review:
 \be
 \ell_s^2 \sim {\ell_p^6 \over V_4}, \qquad g_s^2 \sim \left( {V_4 \over \ell_p^4}\right)^{3/2}. 
 \ee 
 In the conventional field theory normalization for the $(2,0)$ theory, the scalars $\phi^I$ have mass dimension $2$ with kinetic terms proportional to, 
 \be\label{scalarkin}
 \int d^6x \, \partial\phi^I \partial \phi^I. 
 \ee
  The scale $\ell_p$ does not appear in the terms that involve $2$ derivatives in the equations of motion, but only appears in terms involving higher derivative interactions. Reduction of~\C{scalarkin}\ on $K3$ would not naturally give the usual string non-linear sigma model without introducing $\ell_p$ via
  \be
  \int d^2x \int_{K3} \, \partial\phi^I \partial \phi^I \quad \Rightarrow \quad {1\over 4\pi\alpha'}\int d^2x \partial\hat\phi^I \partial \hat\phi^I, \qquad \hat\phi^I =  2\sqrt{\pi} \ell_p^3 \phi^I. 
  \ee
 The redefinition of $\phi^I$ does not involve $V_4$ at all. In a standard field theory reduction, one would have arrived at dimensionless scalars  $\tilde\phi^I = \sqrt{V_4} \phi^I$. This is a caveat in comparing the natural reduction of the $(2,0)$ theory with expectations from a dual heterotic description.

 Let us consider the case of $\M_8=K3\times K3$. With different choices of metric and flux, we find a dual heterotic description consisting of a torsional geometry. The geometry is the fibration of $T^3$ over $K3$ discussed earlier, with the choice of connections $A_i$ appearing in~\C{torsionalmetric} determined by the M-theory flux. The fundamental heterotic string is described by a world-sheet theory with this geometry as a target space. Wrapping an M5-brane on either $K3$ of $\M_8$ gives a physical heterotic string for this background. The two choices of wrapping submanifold correspond to dual heterotic descriptions~\cite{Sethi:2007bw, Becker:2007ea}. 
 
 We can therefore conclude that in this special case of M-theory/heterotic dual pairs, the wrapped M5-brane gives a $D=2$ string with a torsional target space geometry. For a given flux, different choices of metric on $K3\times K3$ can preserve differing amounts of supersymmetry. Different choices of metric also give different numbers of space-time massless modes. We therefore already find a large class of interacting two-dimensional theories from activating the flux degree of freedom. There is no reason at all to restrict to this special case; we can wrap an M5-brane or multiple M5-branes on any $4$-submanifold of $\M_8$ which solves the supergravity equations of motion, subject to mild conditions on the flux. With this motivation, we now turn to the question of describing the couplings on an M5-brane in a general background of $11$-dimensional supergravity.

\section{Coupling an M5-brane to $11$-dimensional Supergravity} \label{hardwork}

The world-volume of the M5-brane is described by its embedding into a target supermanifold, $Z: \Sigma_{6|0} \to \mathcal{M}_{11|32}$. Choosing coordinates on $\mathcal{M}$, we can write $Z$ as
\be
Z^A(\sigma) = \left( X^M(\sigma), \Theta^a(\sigma) \right),
\ee
where $\sigma^i$ are coordinates on $\Sigma$. In addition to the embedding, the world-volume supports a two-form gauge potential $B_2 (\sigma)$. Viewed as coordinates on the supermanifold $\mathcal{M}_{11|32}$, $X$ and $\Theta$ are naturally assigned mass dimension $-1$ and $-1/2$, respectively.  $B_2$ has mass dimension $-1$. This differs from the standard field theory dimensions, which would result from absorbing powers of the Planck length, $\ell_p$, into these fields. We will not perform this renormalization, however, for the discussion of this section.

Before proceeding, we should make a few remarks regarding our assumptions and our reliance on the example scenarios described in section~\ref{fluxbackgrounds}. Throughout the following discussion, we will only assume that the $11$-dimensional background admits at least one Killing spinor and that the M5-brane is Lorentzian. Beginning in section~\ref{cycles}, for reasons described in that section, we will assume that the $G_4$ flux vanishes when pulled-back to the classical world-volume. Finally, only in section~\ref{cycles}, where we describe some specific examples, will we return to the backgrounds described in section~\ref{fluxbackgrounds}.

Let us list our index conventions. M5-brane world-volume coordinates will be denoted $i, j, \ldots$. The $11$-dimensional superspace coordinates will have $A, B, \ldots$ indices which decompose into $11$-dimensional bosonic coordinates $M, N, \ldots,$ and spinor indices $a, b, \ldots$. We further split the $11$-dimensional coordinates into those tangent and transverse to the M5-brane: $X^M = (X^\mu, X^I)$ where $\mu = 0, 1, \ldots 5$ denote the tangent directions. 

Note that $X^\mu$ is not logically identified with the world-volume coordinate $\sigma^i$ before choosing static gauge. This notational distinction will be necessary for clarity in our later discussion. At times, we will also need to denote orthonormal frame indices, which we do by underlining the respective coordinate indices: $\ul{i}, \ul{M}$, etc. For more details regarding our spinor and gamma matrix conventions, see appendix~\ref{GammaMatrices}.

The action which describes the self-interaction of the M5-brane is complicated by the presence of a self-dual three-form, nonlinearly related to the field strength of $B_2$. One strategy to overcome this complication---the PST formulation~\cite{Pasti:1997gx,Bandos:1997ui}---involves the introduction of an auxiliary scalar field $a(\sigma)$ whose equation of motion is responsible for ensuring self-duality. The action is\footnote{The tension of the M5-brane, $T_{M5}$ is $2\pi\left(2\pi\ell_p\right)^{-6}$}
\be\label{PSTAction}
\bes
S_{M5} &= T_{M5}\int d^6 \sigma \left( \sqrt{-\det\left( g_{i j} + \wt{H}_{ i j} \right)} + {\sqrt{-\det g} \over 4} \wt{H}_{i j}{H}^{ij} \right) \cr
&+ T_{M5} \int Z^*C_6 + {1\over 2} \mathcal{H}_3 \wedge Z^*C_3.
\ees
\ee
The metric $g_{ij}$ is the induced metric on the world-volume:
\be
g_{ij } = \partial_i Z^A\partial_j Z^B E_A^{\phantom{A}\ul{M}}E_B^{\phantom{N}\ul{N}}\eta_{\ul{M}\ul{N}},
\ee
with $E_A{}^{\ul{A}}(Z)$ the supervielbeine of the $11$-dimensional background. The auxiliary scalar only couples via its normalized derivative,
\be
v_i = {\partial_i a \over \sqrt{ -g^{jk}\partial_j a\partial_k a}},
\ee
which shows up in the definition of $\wt{H}_{ij}$:
\be
\bes
\wt{H}_{ij} &= v^k \left( \ast \mathcal{H} \right)_{ijk}, \cr
H_{ij} & = v^k \mathcal{H}_{ijk}, \cr
\mathcal{H}_3 &= dB_2 - Z^*C_3.
\ees
\ee

As an effective field theory describing the interactions of the brane degrees of freedom, the PST action includes an infinite set of irrelevant deformations suppressed by powers of $\ell_p$, using the field theory normalization in which the two derivative terms are independent of $\ell_p$. In a trivial Minkowski supergravity background, truncating the PST action to the two derivative terms yields the superconformal action for a single, abelian $(2,0)$ tensor multiplet~\cite{Claus:1997cq}.

In the presence of a non-trivial supergravity background, the action for the tensor multiplet includes couplings to the background fields, like the case of topological twisting where couplings to the connection on the bulk tangent bundle act as background $R$-symmetry gauge fields in the field theory. Supergravity fluxes induce yet more field theory couplings. For example, under the simplifying assumption that the $11$-dimensional $G_4$ flux vanishes on pull-back to the classical brane configuration,\footnote{Also assumed is that the background value of the world-volume three-form $\mathcal{H}_3$ is zero. This is consistent with the Bianchi identity for $\mathcal{H}_3$ only if the previously mentioned assumption about $G_4$ holds.} the quadratic action for the world-volume fermions takes an especially simple form~\cite{Tsimpis:2007sx}, schematically:
\be\label{squad}
S_{\text{quad}} = \int d^6\sigma \sqrt{-\det g} \left(\ov{\Theta} \hat{\gamma}^i \nabla^{(G)}_i \Theta\right),
\ee
where $\nabla^{(G)}$ includes both the aforementioned couplings to the bulk connection as well as couplings to the bulk flux. Its precise form will be given later.

Our primary interest will be in the supersymmetries enjoyed by theories like the one above, when supplemented with the bosons. These supersymmetries descend from the superisometries of the background and are, in general, a subset of them. This is what we will describe next.

\subsection{Symmetries of the M5-brane action}
The M5-brane action enjoys both global and local symmetries. Among the global symmetries are the superisometries\footnote{We actually will be concerned with the supergauge transformations~\cite{Wess:1992cp}. These are described further in appendix~\ref{thetaexpansions}.} of the $11$-dimensional background, each of which is generated by a super-Killing vector $K^A(Z)$ and acts on the $11$-dimensional coordinates as
\be
\delta_K Z^A = K^A.
\ee
The superisometries include the bosonic isometries of the background as well as the fermionic symmetries generated by the Killing spinors $\epsilon$.

Note that in terms of the world-volume fields $X^M$ and $\Theta^a$, these symmetries are generally nonlinearly realized, i.e.\ they are spontaneously broken symmetries. This will also apply to a subset of the local symmetries; thus, finding manifestly supersymmetric brane configurations is equivalent to finding backgrounds for which a subset of these nonlinearly realized symmetries are actually linearly realized when restricted to the physical degrees of freedom. We will always assume that the background admits odd superisometries, which we define as having a nonzero $K^a\big|_{\Theta =0}$, though we will later make some comments on what happens when this is not true.

The local symmetries include world-volume diffeomorphisms and the fermionic kappa-symmetry. The latter act on the embedding and world-volume fields as follows, 
\be
\bes
\delta_\kappa Z^A &= \Delta^A, \cr
\delta_\kappa B_2 & = Z^* \iota_{\Delta} C_3, \cr
\Delta^{A} &= \left( \left(1 + \Gamma_\kappa \right) \kappa \right)^{\ul{a}} E_{\ul{a}}{}^A,
\ees
\ee
where $\kappa(\sigma)$ is an arbitrary, Majorana spinor and $E_{\ul{A}}{}^A$ are the inverse supervielbeine. The matrix $\Gamma_\kappa$ satisfies
\be
\text{tr} \left( \Gamma_\kappa \right)= 0, \quad \Gamma_\kappa^2 = 1,
\ee
and so $(1+ \Gamma_\kappa)$ acts as a projector, leaving only $16$ independent components in the kappa transformation. These surviving kappa transformations act to render half of the components of $\Theta$ unphysical. This agrees with the field theory multiplet structure  since the tensor multiplet contains $16$ and not $32$ fermionic degrees of freedom.

Now we must ask, when does a brane configuration possess manifest supersymmetries on its world-volume? As in the case of supergravity, a bosonic field configuration, schematically denoted $\mathcal{B}$, is supersymmetric if a fermionic symmetry transformation preserves its bosonic nature. In other words, a symmetry generated by the fermionic parameter $\epsilon$ is a supersymmetry if the variation of each fermion vanishes:
\be\label{shiftsusy}
\delta_\epsilon \psi \Big|_{\mathcal{B}} = 0,
\ee
for all fermionic fields $\psi$ in the theory. On the brane, the only fermions are $\Theta$. Furthermore, only those fermionic symmetries that act on $\Theta$ by a $\Theta$-independent shift could possibly violate~\C{shiftsusy}. These symmetries only include the superisometries and the kappa symmetry. Therefore, we require:
\be
\delta \Theta\Big|_{\mathcal{B}} = \left(1 + \Gamma_\kappa\Big|_{\mathcal{B}}\right) \kappa + K\Big|_{\mathcal{B}} = 0.
\ee
Projecting onto the subspace unaffected by kappa symmetry yields the supersymmetric cycle condition, also called the calibration condition, of~\cite{Becker:1995kb}:\footnote{We have assumed that the M5-brane world-volume is Lorentzian. If it is Euclidean, then this equation is modified by factors of $i$.}
\be\label{susycycle}
\left(1 - \Gamma_\kappa \right)K\Big|_{\mathcal{B}} =0.
\ee
The matrix $\Gamma_\kappa$ contains information on the embedding and the background fields, so this puts a constraint on which superisometries are respected by the brane plus background configuration. Or, put differently, the supersymmetric cycle condition tells us which brane embeddings in a given background preserve some supersymmetry.

\subsection{Supersymmetric cycles}\label{cycles}
What can we learn from~(\ref{susycycle}) about the supersymmetric embeddings of M5-branes? It turns out that submanifolds satisfying this condition are described by calibrated---or generalized calibrated---geometries~\cite{Harvey:1982xk,Gutowski:1999tu}. There is a large literature on these topics, and it will be enough for our purposes to point out that solutions to~(\ref{susycycle}) do exist and to discuss some of the properties of a few examples.

For now, let us restrict to the scenario described in section~\ref{fluxbackgrounds}, which is that 
\be
\mathcal{M}_{11} = \mathbb{R}^{2,1} \times_W \mathcal{M}_8,
\ee
with $\mathcal{M}_8$ conformally a Calabi-Yau $4$-fold. Also, the M5-brane wraps a four-cycle $\wt{\Sigma}_4 \subset \mathcal{M}_8$. In this case, the Killing spinors of $\mathcal{M}_8$ are conformally related to those of the $CY_4$~\cite{Becker:1996gj}. Furthermore, we will make the simplifying assumption that the pull-back of the $4$-form flux to the brane world-volume, evaluated in the classical configuration, vanishes. This allows us to set the background value of $\mathcal{H}_3$ to zero, which we will do. Under this last assumption, $\Gamma_\kappa$ takes a very simple form:
\be
\Gamma_\kappa\Big|_{\mathcal{B}} = \Gamma_{\ul{0}} \cdots \Gamma_{\ul{5}} \coloneqq \wt{\Gamma},
\ee
where the indices $\ul{0} \ldots \ul{5}$ run over the directions of the world-volume in the orthonormal frame. Our conventions are described in appendix~\ref{GammaMatrices}. This is the same form that $\Gamma_\kappa$ takes in the absence of flux. It is the $11$-dimensional gamma matrix that measures chirality along the world-volume, and the supersymmetric cycle condition reduces to the requirement that $K$ has positive world-volume chirality.

We claim that in this case, the supersymmetric cycles of $\mathcal{M}_8$ are simply related to those of the underlying $CY_4$ to which $\mathcal{M}_8$ is conformally related. Let us motivate this with an example. Suppose $\Sigma_4 \subset CY_4$ is a K{\"a}hler cycle, i.e. a submanifold whose volume form is given by the restriction to $\Sigma_4$ of the square of the K{\"a}hler form $J$:
\be
\text{Vol}(\Sigma_4) = J \wedge J \Big|_{\Sigma_4}.
\ee
This follows from~(\ref{susycycle}) after multiplying that equation by the dual spinor and integrating over $\Sigma_4$ while further relating $J$ to a Killing spinor bilinear~\cite{Becker:1995kb}.

When we include flux, we can take the same underlying topological cycle as $\Sigma_4$, but in the new metric on $\mathcal{M}_8$, which is conformally related to the $CY_4$ metric, the volume form of this new cycle has changed by a conformal factor; we'll denote the new cycle by $\wt{\Sigma}_4$.  On the other hand, if we use the Killing spinor to construct a bilinear $\wt{J}$, which is conformally related to the previous K\"ahler form $J$ (and called a generalized K\"ahler form), then keeping track of the conformal factors we can show that
\be
\operatorname{Vol}(\wt{\Sigma}_4)=\wt{J}\wedge\wt{J}\,\Big|_{\wt{\Sigma}_4}\propto J\wedge J\Big|_{\wt{\Sigma}_4}.
\ee
When we have this condition, we can call $\wt{\Sigma}_4$ a generalized K\"ahler four-cycle on $\mathcal{M}_8$.  Moreover, if we again multiply (\ref{susycycle}) by the dual spinor, now in the presence of fluxes, we see that this generalized calibration condition is exactly what we need to preserve supersymmetry on the brane wrapping $\wt{\Sigma}_4$.  Analogously, we can show that calibrated K\"ahler cycles of different dimensions or calibrated special Lagrangian cycles of the $CY_4$ also lead to appropriate supersymmetric generalized calibrated cycles on the conformal Calabi-Yau $\mathcal{M}_8$ in the presence of flux.  Similar statements should hold in flux backgrounds whenever the $G$-structures are only modified by a conformal factor.

With this relation between the cycles of $\mathcal{M}_8$ and $CY_4$, we can characterize the supersymmetric M5-brane embeddings with flux by the corresponding embeddings without flux, but with the added requirement that the flux satisfy our assumption that it vanish on the classical world-volume. To give some examples of the two-dimensional field theories that can result from our construction, we describe the cycles and fluxes that satisfy our assumptions below.

Starting with a generic Calabi-Yau $4$-fold, which we take to have holonomy $SU(4)$ and not a proper subgroup, the supersymmetric $4$-cycles have been studied and classified.\footnote{See, for example,~\cite{Gauntlett:2003di}.} We will focus on two types: K{\"a}hler and special Lagrangian. In the absence of flux, an M5-brane wrapping a K{\"a}hler cycle inside of a $CY_4$ yields a two-dimensional theory with $\mathcal{N} = (0,2)$ supersymmetry. Alternatively, a special Lagrangian cycle yields $\mathcal{N} = (1,1)$. There is a third possibility, which is to wrap a Cayley submanifold~\cite{Becker:1996ay}. In this case, the resulting theory has $\mathcal{N} = (0,1)$. 

As described in section \ref{fluxbackgrounds}, the possible supersymmetric fluxes can preserve either all or half of the supersymmetries of the metric-only background. However, the flux that preserves two supersymmetries does not satisfy our assumption regarding its pull-back to the world-volume. As a reminder, this component of the flux is
\be
J\wedge J + {3\over 2} {\rm Re}(\Omega). \nonumber
\ee
For either a K{\"a}hler or special Lagrangian cycle, this flux has a non-vanishing pull-back to the world-volume. Therefore, under our assumptions, we can only include the primitive $(2,2)$ flux.

If the $CY_4$ is actually $K3 \times \wt{K3}$ and $\Sigma_4 \simeq K3$, the resulting two-dimensional theory has either $\mathcal{N} = (0,4)$ or $\mathcal{N} = (2,2)$ supersymmetry, without accounting for flux. As before, it is possible for fluxes to partially break the supersymmetry of the pure metric background~\cite{Dasgupta:1999ss}. In this case, they can preserve all, half, or one quarter of the supersymmetries. The most general, supersymmetric flux on a $K3 \times \wt{K3}$ appears in~\cite{Prins:2013wza}:
\be
\bes  
G_4 &= c j \wedge \tilde{\jmath} + a~\textrm{Re}\left( \omega \wedge \wt{\omega} \right) + \textrm{Re}\left( b \omega^* \wedge \wt{\omega} \right) \cr
&\phantom{=}+ (4a - 2c)\left( \textrm{vol}_4 + \wt{\textrm{vol}}_4 \right) + f^{\alpha \beta} l_\alpha \wedge \wt{l}_{\beta} \cr
&\phantom{=}+ h\omega \wedge  \tilde{\jmath}  + h^* \wt{\omega} \wedge j + \textrm{c.c.}
\ees
\ee
The constants $a, c, f^{\alpha \beta} \in \mathbb{R}$ and $b, h \in \mathbb{C}$ but are otherwise unconstrained. The forms $j$, $\textrm{Re}~ \omega$, and $\textrm{Im}~ \omega$ comprise a triplet of complex structures on $K3$. The forms $l_\alpha$ are a basis of anti-self-dual two-forms. Let us restrict to the case of  $\mathcal{N} = (0,4)$ supersymmetry. To satisfy our assumptions regarding the pull-back of the flux, we must have $4a-2c = 0$, i.e.\
\be
\bes
G_4 &= 2a j \wedge  \tilde{\jmath}  + a~\textrm{Re}\left( \omega \wedge \wt{\omega} \right) + \textrm{Re}\left( b \omega^* \wedge \wt{\omega} \right) \cr
&\phantom{=}+  f^{\alpha \beta} l_\alpha \wedge \wt{l}_{\beta} + h\omega \wedge  \tilde{\jmath}  + h^* \wt{\omega} \wedge j + \textrm{c.c.}
\ees
\ee
At generic points on the locus $a = h =0$, the flux preserves only half of the background supersymmetries. Off this locus, it preserves only a quarter. To preserve all of the background supersymmetries, it is necessary for the flux to be invariant under a nontrivial subgroup of the $SU(2) \times \wt{SU(2)}$ rotations acting on $\{j, \textrm{Re}~ \omega,\textrm{Im} ~\omega\}$ and $\{ \tilde{\jmath}, \textrm{Re}~ \wt{\omega},\textrm{Im} ~\wt{\omega}\}$. This is only possible if $a=b=h=0$.

In summary, wrapping an M5-brane on a four-cycle inside of a Calabi-Yau $4$-fold in the presence of flux can lead to two-dimensional field theories with $\mathcal{N} = (1,1), (0,2)$ and $\mathcal{N} =(0,1)$. 
The remaining possibility is the strangest; namely, to turn on a $G_4$ flux with a $(4,0)$ component so that all the ambient supersymmetries are broken. We might expect the resulting $D=2$ string to be completely non-supersymmetric since there are no ambient superisometries. However, if we were wrapping a $K3$ submanifold of, for example, $K3\times K3$ then we should have constructed the physical heterotic string. The world-sheet heterotic string needs $(0,1)$ supersymmetry regardless of whether the background possesses any space-time supersymmetry. In addition, $(0,1)$ is a very weak condition on the resulting sigma model. It is, therefore, possible that the $D=2$ theory is actually $(0,1)$ supersymmetric even if the flux breaks all the bulk supersymmetries. 

\subsection{World-volume supersymmetry}
Next, we come to the question of determining the supersymmetry transformations of the physical fields on the world-volume. First, what are the physical fields on the world-volume? We saw before that the kappa-symmetry renders only half of the $\Theta$ variables physical. We will fix the gauge so as to make this manifest. Additionally, the world-volume diffeomorphisms allow us to remove $6$ of the $11$ scalar degrees of freedom. Ultimately, the bosons $\phi^I$ are a section of the normal bundle to $\Sigma_6$ in $\mathcal{M}_{11}$ and the fermions $\Theta$ are a (chiral) section of the spin bundle associated to the pulled-back tangent bundle of the embedding $Z$:
\be
\bes
\phi &\in \Gamma\left[\mathcal{N}\right], \cr
\Theta &\in \Gamma\left[ \mathcal{S}^-\left(T\Sigma\right) \otimes \mathcal{S}\left(\mathcal{N}\right) \right].
\ees
\ee

To make this more precise, we fix the kappa-symmetry gauge. Define the projection matrices
\be
\left( \mathcal{P}_{\pm} \right)^a{}_b = {1 \over 2} \left( 1 \pm \wt{\Gamma} \right)^a{}_b,
\ee
which project $11$-dimensional spinors onto the subspaces with positive or negative chirality, defined with respect to the world-volume $\Sigma_6$. Our gauge choice is
\be
\Theta_+ \coloneqq \mathcal{P}_+ \Theta = 0.
\ee
Generically, kappa-transformations and superisometries move away from this gauge slice. The subset which preserves the gauge choice satisfies
\be\label{compensatingtransf}
\mathcal{P}_+\Big( (1 + \Gamma_\kappa)\kappa + K \Big) = 0.
\ee 
This constraint determines the kappa-transformation, $\kappa(K)$, which combines with $K$ to preserve the gauge choice. Together, their action on $\mathcal{P}_-\Theta \coloneqq  \Theta_-$ give the supersymmetry transformations of $\Theta_-$:
\be
\delta \Theta_- = \mathcal{P}_- \Big( (1 + \Gamma_\kappa) \kappa + K \Big).
\ee
Only the subset of these transformations also satisfying~(\ref{susycycle})\ will be linearly realized.

To get a more concrete handle on these transformations, we decompose $\Gamma_\kappa$ in the chiral basis
\be
\wt{\Gamma} = \begin{pmatrix} 1 & 0 \\ 0 & -1 \end{pmatrix}.
\ee
To satisfy $\rm{tr} \left( \Gamma_\kappa \right) =0$ and $\Gamma_\kappa^2 = 1$, $\Gamma_\kappa$  must take the form~\cite{Kallosh:1997ky}
\be
\Gamma_\kappa = \begin{pmatrix} B & (1-B^2)A^{-1} \\ A & -B \end{pmatrix},
\ee
with $A$ and $B$ commuting. Then,~(\ref{compensatingtransf}) is solved by
\be
\kappa_+ = -(1+B)^{-1}K_+ - (1-B)A^{-1} \kappa_-.
\ee
The transformations of $\Theta_-$ are then given by,
\be\label{exactthetatransf}
\delta\Theta_- = -(1+B)^{-1}AK_+ + K_-.
\ee
Note that the dependence on $\kappa_-$ dropped out. We should also include a $K$-dependent world-volume diffeomorphism, whose role is to fix the gauge associated to the choice of scalar degrees of freedom; however, it will play no role in our analysis. Ignoring this diffeomorphism, the above expression is exact. We will be interested in expanding~\C{exactthetatransf}\ in $\Theta$ to obtain the leading transformations. In appendix~\ref{thetaexpansions}, we obtain the following expansion of the superisometry:
\be \label{expsuperisom}
K^M = {i \over 2} \ov{\Theta}_- \Gamma^M \epsilon, \qquad K^a = \epsilon^a,
\ee
where $\epsilon$ solves the $11$-dimensional gravitino variation condition, so it is a Killing spinor of the spacetime. Using~\C{expsuperisom}, the leading-in-$\Theta$ transformation is
\be
\delta\Theta_- = -(1+B)^{-1}A\epsilon_+ + \epsilon_-.
\ee
We see that the supersymmetries associated to $\epsilon_-$ are nonlinearly realized. In other words, the linear supersymmetries are chiral, as they are for the $(2,0)$ theory in a trivial background.

Returning to the bosons, the transformations of the $\phi^I$ to leading order in a $\Theta$-expansion are
\be
\bes
\delta \phi^I &=  i e_{\ul{I}}{}^I(\sigma,\phi) \ov{\Theta}_- \Gamma^{\ul{I}} \epsilon_+ + {i \over 2} e_{\ul{\mu}}{}^I(\sigma,\phi) \ov{\Theta}_- \Gamma^{\ul{\mu}}(1+B)^{-1} A \epsilon_+ \cr
&\phantom{=} + {i \over 2} e_{\ul{\mu}}{}^I(\sigma,\phi) \ov{\Theta}_- \Gamma^{\ul{\mu}} \epsilon_- ,
\ees
\ee
where we have used the $\Theta$-expansions of the supervielbeine and $K^A$ given in appendix~\ref{thetaexpansions}. We have retained the $\phi$-dependent bulk vielbeine $e$ to emphasize that we have not truncated to any order in $\phi$. However, we have used the gauge freedom of world-volume diffeomorphisms and bulk frame rotations to simplify the expressions a little. Specifically, coordinates on $\mathcal{M}_{11}$ are chosen so that the world-volume $\Sigma_6$ is isomorphic to the submanifold $\phi =0$. The bulk metric restricted to this submanifold is block diagonal, and we can also choose a framing at $\phi = 0$ which is block diagonal. So, in a $\phi$ expansion, only the first term would survive at leading order.

At leading order in $\Theta$, the $2$-form has the transformation
\be
\delta B_{ij} = -i e_{i}{}^{\ul{M}} e_{j}{}^{\ul{N}} \ov{\Theta}_-  \Gamma_{\ul{MN}} (1 + \Gamma_\kappa )\kappa(\epsilon),
\ee
where we have defined $e_i{}^{\ul{M}} = (X^*e^{\ul{M}})_i$ as the pull-back of the bulk vielbeine using the bosonic embedding $X$. At $\phi = 0$, $e_i{}^{\ul{M}}$  reduce to the world-volume vielbeine.

After all of this discussion, we have yet to specify $\Gamma_\kappa$ and, hence, $A$ and $B$. Let us remedy this. The kappa symmetry projector for the M5-brane appears in~\cite{Bandos:1997ui}:
\be
\Gamma_\kappa = {1\over \sqrt{\det(1 + \tilde{H} )}} \left( \wt{\gamma} + {1 \over 2}v^i \wt{H}^{j k} \gamma_{ijk} - {1 \over 8} \epsilon^{i_1 \ldots i_5 j}\wt{H}_{i_1 i_2}\wt{H}_{i_3 i_4}v_{i_5}v^i\gamma_{i j} \right).
\ee
The matrices $\gamma_i$ are the pull-backs of the bulk gamma matrices. Specifically,
\be
\gamma_i = \partial_i Z^A E_A{}^{\ul{M}} \Gamma_{\ul{M}}.
\ee
These matrices furnish a representation of the six-dimensional Clifford algebra,
\be
\acom{\gamma_i}{\gamma_j} = 2g_{ij},
\ee
and the associated chirality matrix is
\be
\wt{\gamma} = {1 \over 6!} \epsilon^{i_1 \ldots i_6} \gamma_{i_1 \ldots i_6}.
\ee
Note that this generally differs from the pull-back of the matrix $\wt{\Gamma}$. Only when $\phi = \Theta = 0$ do they agree.

Without employing some expansion, there is little more we can glean about the world-volume supersymmetry from these expressions.\ $A$ and $B$ are the off- and on-diagonal components, respectively, and they have a complicated structure. In a $\Theta$ expansion of the supersymmetries, $\Gamma_\kappa$ contributes only bosonic terms at leading order, as can be seen from the discussion above. This doesn't necessarily make the expressions simpler, though, because the bosonic $\Gamma_\kappa$ still has a complicated form.

To get a more concrete handle on $\Gamma_\kappa$, we will use a world-volume momentum expansion. This is the same expansion used in a similar context in~\cite{Harvey:1999as}. In this expansion, we assign degree $1$ to derivatives with respect to world-volume coordinates and degree $0$ to the world-volume bosons, as well as to derivatives of bulk objects with respect to the normal coordinates. Additionally, the world-volume fermions are assigned degree $1/2$, while the supersymmetry parameter $\epsilon$ has degree $-1/2$. Thus,
\be
[\partial_i \phi^I ] = [\mathcal{H}_3] = 1, \qquad [\Theta_-] = 1/2.
\ee
This expansion is often employed in the expansion of supersymmetric effective field theories. The leading supersymmetry transformations preserve this notion of degree, while higher-order transformations raise the degree.

Generally, this degree expansion is inconsistent with our truncation of the supersymmetry transformations to leading order in $\Theta$.
Specifically in the action of supersymmetry on $\Theta_-$, there could in principle be terms of the form $\mathcal{O}(\epsilon \Theta^2)$ as long as the coefficient has degree zero. The coefficients are typically bulk objects like fluxes or connections, and their degree assignment requires some care. 

In~\cite{Harvey:1999as}, it was argued that it is consistent to assign degree 1 to all geometric objects unless they are associated with a subspace of $\mathcal{M}_{11}$ that carries a product structure. To be more specific, suppose $\mathcal{M}_{11}$ decomposes as
\be
\mathcal{M}_{11} = \mathcal{M}_{6+p} \times \mathcal{M}_{5-p}, \nonumber
\ee
where $\Sigma_6$ is a submanifold of $\mathcal{M}_{6+p}$. 
In general bulk objects, like connections, that depend on the coordinates of $\mathcal{M}_{5-p}$ and derivatives with respect to those coordinates can be assigned degree 0. This assignment cannot be made consistently for objects with indices or derivatives involving $\mathcal{M}_{6+p}$. For example, the formulae for relating Christoffel symbols to spin connections can mix derivative type, and therefore degree.

Similar care must be taken to define the degree of the $4$-form flux, when there is a non-zero $G_4$ on the world-volume. This is because the components of $C_3$ on the world-volume mix with $dB_2$ and so must be given degree 1. On the other hand, the $4$-form seems to appear along with connections in various formulae, and we would also like to assign the flux degree 1. These two considerations are in tension. However, if we assume that there is no $4$-form along the world-volume, there are no inconsistencies associated with assigning degree 1 to $G_4$.  This has been our working assumption for several reasons, as we discussed above eq.~\C{squad}. 

With this prescription for the degree expansion, it is simple to include the $\mathcal{O}(\epsilon\Theta^2)$ terms that may contribute to the $\Theta$ transformation. Returning to the exact expression~(\ref{exactthetatransf}), the only important contribution will come from $K_-$. This is because, as we will see, there is no degree $0$ piece to $A$. Since we are interested in the linearly realized supersymmetries, which are parameterized by $\epsilon_+$, it is simple to deduce from~(\ref{superkilling}) which parts of $K_-$ are important. They are
\be
{i \over 8}\omega_{\ul{I \mu \nu}} \left( \ov{\Theta}_- \Gamma^{\ul{I}} \epsilon_+\right) \left( \Gamma^{\ul{\mu \nu}} \Theta_- \right) + {i \over 8}\omega_{\ul{IJK}} \left( \ov{\Theta}_- \Gamma^{\ul{I}} \epsilon_+\right) \left( \Gamma^{\ul{JK}} \Theta_- \right). \nonumber
\ee 
It is a drastic simplification that these terms do not depend on $G_4$.

To obtain the supersymmetries to leading order in this expansion, we need the degree expansions of the following quantities:
\be
e_i{}^{\ul{M}}, \quad \mathcal{H}_{ijk}. \nonumber
\ee
The degree counting of the frame is subtle for the aforementioned reasons. We have also yet to (partially) fix the local Lorentz symmetry. This is accomplished by choosing $e_I{}^{\ul{\mu}}(\sigma, \phi) = 0$, which leaves manifest only the $SO(5,1)\times SO(5)$ subgroup of $SO(10,1)$ rotations on the frame. This choice ensures that the other off-diagonal component of the vielbeine, $e_\mu{}^{\ul{I}}$, has no degree 0 part. Its first non-vanishing component is related to the connection on the normal bundle, which is of degree 1. 

Furthermore, expanding this component of the vielbeine as well as $e_\mu{}^{\ul{\mu}}$ in degree is the same as expanding in $\phi$, because higher-order in $\phi$ contributions have higher degree since they are related to curvatures on the world-volume. So, in a leading degree expansion, we are only forced to maintain the $\phi$-dependence of $e_I{}^{\ul{I}}$, and this is only necessary when the complement of $\Sigma_6$ in $\mathcal{M}_{11}$ has the product the structure mentioned before. Note that these details are important for us since they directly affect one of our primary examples, $\M_8= K3 \times K3$. 
 
The three-form expansion is more straightforward. All of the components of $G_4$ are of degree 1, so in the degree expansion of the three-form, we must retain all of the $\phi$-dependence. To first order in degree:
\be
\mathcal{H}_{ijk} = \left(dB\right)_{ijk} + \phi^IG_{Iijk}(\phi) + 3\partial_{[i}\left( \phi^I C_{jk]I} \right).
\ee
The last term is an exact form on the world-volume. We can fix the bulk gauge symmetry under which $B_2$ shifts by a two-form to remove this piece, leaving
\be
\mathcal{H}_{ijk} = \left(dB\right)_{ijk} + \phi^IG_{Iijk}.
\ee
Assembling the above results, we find the leading supersymmetries on the world-volume:
\be\label{susy}
\bes
\delta \phi^I &= -i e_{\ul{I}}{}^I(\phi) \left(\ov{\epsilon}_+ \Gamma^{\ul{I}} \Theta_-\right) , \cr
\delta B_{ij} &= -i\ov{\epsilon}_+ \hat{\gamma}_{ij} \Theta_-, \cr
\delta \Theta_- &= -{1 \over 2} e_I{}^{\ul{I}}(\phi) \left(\nabla_i \phi^I \hat{\gamma}^i \Gamma_{\ul{I}} \epsilon_+\right)  -{1\over 24} \mathcal{H}_{ijk}\hat{\gamma}^{ijk} \epsilon_+ \cr
&\phantom{= -} + {i \over 8}\omega_{I \ul{ \mu \nu}} e_{\ul{I}}{}^I (\phi) \left( \ov{\Theta}_- \Gamma^{\ul{I}} \epsilon_+\right) \left( \Gamma^{\ul{\mu \nu}} \Theta_- \right) + {i \over 8}\omega_{I\ul{JK}} e_{\ul{I}}{}^I (\phi)\left( \ov{\Theta}_- \Gamma^{\ul{I}} \epsilon_+\right) \left( \Gamma^{\ul{JK}} \Theta_- \right).
\ees
\ee
The derivative $\nabla$ is the covariant derivative on the combined normal and tangent bundles.
The matrices $\hat{\gamma}$ are the pull-backs of $\Gamma$ to the classical world-volume. Specifically, they are the previously defined $\gamma$ matrices, evaluated at $\phi = \Theta = 0$. We have included terms in the last line that may be excluded if the space does not have a product structure, as mentioned earlier in this discussion. Also, in such a case, the inverse vielbeine, $e_{\ul{I}}{}^I(\phi)$, should be replaced with their ``hatted'' versions, where ``hatted" means evaluated at $\phi=0$. 

\subsection{Closure and equations of motion}
We can use the closure of the supersymmetry algebra to derive equations of motion for the world-volume fields. Specifically, we demand that the commutator of two supersymmetry transformations acting on any field yields a bosonic symmetry action on that field, when we restrict to on-shell configurations. For simplicity, we will restrict our discussion to the supersymmetry transformations that are linear in $\phi$ and $\Theta$.

Acting on the two-form, we find that supersymmetry closes onto isometries and gauge transformations only when the field strength of $B_{ij}$ is no longer self-dual. Specifically,
\be\label{threeformEOM}
(dB)_{ijk}^- = -\phi^IG_{I ijk}^- \quad \Rightarrow \quad {\mathcal H}_{ijk}^-=0.
\ee
To derive this result, it is necessary to make use of a particular constraint on the fluxes that results from the existence of a Killing spinor. This constraint says that the bulk fluxes, when restricted to $\phi = 0$, i.e. the brane, satisfy
\be\label{Gconstraint}
G_{ijk \ul{I}} \hat{\gamma}^{ijk} \epsilon_+ = G_{i \ul{JKL}} \hat{\gamma}^i \Gamma_{\ul{I}}{}^{\ul{JKL}} \epsilon_+.
\ee

The constraints of closure yield no new information when acting on the scalar fields. Acting on the fermion fields, we can find the fermion equation of motion. However, it is easier to find the fermion equation of motion by acting with supersymmetry on the equation of motion for $\mathcal{H}_{ijk}^-$, given in~(\ref{threeformEOM}). In this case, we find the result:
\be
\hat{\gamma}^i \nabla_i^{(G)} \Theta_- = 0,
\ee
where the fluxed covariant derivative acting on $\Theta_-$ is defined to be
\be\label{fermioneom}
\hat{\gamma}^i\nabla_i \Theta_- -{1 \over 48}G_{i \ul{KLM}}\epsilon^{\ul{KLM IJ}} \hat{\gamma}^i \Gamma_{\ul{IJ}} \Theta_- +{1\over 24} G_{ijk \ul{I}} \hat{\gamma}^{ijk}\Gamma^{\ul{I}} \Theta_-.
\ee
This equation of motion agrees in qualitative form with the result found in~\cite{Kallosh:2005yu}. 

The similarity of our supersymmetry variations and equations of motion to those listed in the $D=6$ conformal supergravity formalism of~\cite{Bergshoeff:1999db}\ is quite interesting. In particular, our supersymmetry variations appearing in eq.~(\ref{susy}) are very closely related to their equations after an identification of their auxiliary fields with our flux degrees of freedom.\footnote{Specifically, the auxiliary tensor they refer to as $T^{ij}_{abc}$ is our $G_{ijk I}$ and their $V_\mu^{ij}$ is a combination of our normal bundle connection and the flux component $G_{i IJK}$.} The equations of motion also match after this identification. This suggests a surprising and fascinating relation between the off-shell formalism for coupling a $D$-dimensional field theory to a rigid $D$-dimensional supersymmetric background and the on-shell coupling of a brane to higher-dimensional supergravity. We hope to explore this relation further.

\section*{Acknowledgements}

It is our pleasure to thank Dongsu Bak, Katrin Becker, Jerome Gauntlett, Greg Moore, Dave Morrison, Ergin Sezgin and Dimitrios Tsimpis for helpful discussions. These results were presented at a number of conferences earlier this year, prior to publication. Specifically, the ``Strings, Branes and Holography" workshop held at the Mitchell Institute at Texas A\&M, the Great Lakes Strings Conference held at the University of Michigan, the ``F-theory at the Interface of Particles and Mathematics'' workshop in Aspen, and the joint KIAS-YITP Workshop on ``Geometry in Gauge Theories and String Theory." S.~S. would like to thank the respective organizers of these workshops for their hospitality. 

This work was performed, in part, at the Aspen Center for Physics, which is supported by National Science Foundation grant PHY-1066293. T.~M. and S.~S. are supported, in part, by NSF Grant No.~PHY-1316960. D.~R. is supported by the grants PHY-1214344 and NSF Focused Research Grant DMS-1159404, and by the George P. and Cynthia W. Mitchell Institute for Fundamental Physics and Astronomy.

\newpage
\appendix
\section{Spinors and Gamma Matrices}\label{GammaMatrices}

Our $11$-dimensional spinors will carry upper indices, so the $\Gamma$ matrices will naturally be of the form $\left(\Gamma^{\ul{M}}\right)^{a}{}_{b}$. These matrices will be field-independent and satisfy the $11$-dimensional Clifford algebra:
\be
\acom{\Gamma^{\ul{M}}}{\Gamma^{\ul{N}}} = 2\eta^{\ul{MN}}.
\ee
In the presence of branes, we will want to group these matrices into those along and transverse to the brane. We will use the indices $\{\mu, \ul{\mu} \}$ and $\{I, \ul{I}\}$ to denote the $\{$coordinate, orthonormal$\}$ directions along and transverse to the brane, respectively. Along the M5-brane, we can define a chirality matrix,
\be
\wt{\Gamma} = \Gamma_{\ul{0} \ldots \ul{5}},
\ee
with $\ul{0} \ldots \ul{5}$ being the world-volume directions.

We will often work in a basis in which $\wt{\Gamma}$ is diagonal. Because of the (anti-)commutation relations
\be
\bes
\acom{\Gamma^{\ul{\mu}}}{\wt{\Gamma}} &= 0, \cr
\com{\Gamma^{\ul{I}}}{\wt{\Gamma}} &=0,
\ees
\ee
the matrices $\Gamma^{\ul{\mu}}$ are off-diagonal, while $\Gamma^{\ul{I}}$ are diagonal in this basis.

Furthermore, in $11$-dimensions, there is an antisymmetric charge conjugation matrix $\mathcal{C}_{ab}$ and its inverse $\mathcal{C}^{ab}$. This matrix acts as an intertwiner between the equivalent representations:
\be
\mathcal{C}\Gamma_{\ul{M}} \mathcal{C}^{-1} = -\Gamma_{\ul{M}}^T.
\ee
$\mathcal{C}$ will be used to raise and lower spinor indices. The antisymmetry of $\mathcal{C}$ leads to the following symmetry properties for bilinears in Majorana fermions:
\be
\ov{\Theta} \Gamma^{\ul{M_1} \ldots \ul{M_k}} \epsilon = \Theta^T \mathcal{C}\Gamma^{\ul{M_1} \ldots \ul{M_k}} \epsilon = 
\begin{cases}
+\ov{\epsilon} \Gamma^{\ul{M_1} \ldots \ul{M_k}} \Theta, &\text{if } k =  0, 3, 4, \\
-\ov{\epsilon} \Gamma^{\ul{M_1} \ldots \ul{M_k}} \Theta, &\text{if } k = 1, 2, 5,
\end{cases}
\ee
where $\Theta$ and $\epsilon$ are taken to be anti-commuting. The above is a particular result of the (anti-)symmetry of the gamma matrices themselves
\be
\left( \mathcal{C}\Gamma^{\ul{M_1} \ldots \ul{M_k}} \right)^T = 
\begin{cases}
- \mathcal{C}\Gamma^{\ul{M_1} \ldots \ul{M_k}}, &\text{if } k =  0, 3, 4, \\
+\mathcal{C}\Gamma^{\ul{M_1} \ldots \ul{M_k}}, &\text{if } k = 1, 2, 5,
\end{cases}
\ee
Furthermore, it will be important to note that with respect to the basis in which $\wt{\Gamma}$ is diagonal, $\mathcal{C}$ is off-diagonal, which follows from
\be
\acom{\mathcal{C}}{\wt{\Gamma}} = 0,
\ee
which itself follows from the definition of $\mathcal{C}$ as an intertwiner. This means that for a spinor of given six-dimensional chirality $\wt{\Gamma} \epsilon = \pm\epsilon$, the charge conjugate spinor has the opposite chirality.

\section{Theta Expansions}\label{thetaexpansions}
The expansion of the bulk supervielbeine in the $\Theta$ coordinates is found in~\cite{Tsimpis:2004gq}. For reference:
\be
\bes
E_M{}^{\ul{M}} &= e_M{}^{\ul{M}} -{i \over 2} \ov{\Theta}\Gamma^{\ul{M}}\omega_{M}\Theta
+ {i\over 2} \ov{\Theta}{\mathcal{T}_M}^{NPQR}\Gamma^{\ul{M}}\Theta G_{NPQR}, \cr
E_{M}{}^{\ul{a}} &=\left(\ov{\Theta}\omega_M\right)^{\ul{a}} -\left(\ov{\Theta} {\mathcal{T}_M}^{NPQR}G_{NPQR}\right)^{\ul{a}}, \cr
E_{a}{}^{\ul{M}} &= -{i\over 2}\left(\Gamma^{\ul{M}}\Theta\right)_a, \cr
E_{a}{}^{\ul{a}} &=\delta_a{}^{\ul{a}}+ {i\over 6}G_{NPQR}{(D_1^{NPQR})_a}^{\ul{a}},
\ees
\ee
with $\omega_M$, ${\mathcal{T}_M}^{NPQR}$, and ${(D_1^{NPQR})_a}^{\ul{a}}$ defined as
\be
\bes
\omega_M &= {1\over 4} \omega_M{}^{\ul{MN}}\Gamma_{\ul{MN}}, \cr
{\mathcal{T}_M}^{NPQR} &= -{1\over 288}\left( \Gamma_M{}^{NPQR} + 8\delta_{M}{}^{[N}\Gamma^{PQR]}\right), \cr
{(D_1^{NPQR})_a}^{\ul{a}} &= \left(\ov{\Theta} \Gamma^M\right)_a\left( \ov{\Theta}{\mathcal{T}_M}^{NPQR}\right)^{\ul{a}} + {1\over 4}\left( \ov{\Theta}{\mathcal{R}_{LM}}^{NPQR}\right)_a\left(\ov{\Theta} \Gamma^{LM}\right)^{\ul{a}}, \cr
{\mathcal{R}_{LM}}^{NPQR} &= {1\over 144} \left( {\Gamma_{LM}}^{NPQR} + 24 {\delta_L}^{[N}{\delta_M}^{P} \Gamma^{QR]} \right).
\ees
\ee
It will at times be more convenient to refer the combination ${\mathcal{T}_M}^{NPQR} G_{NPQR}$ as $G_M$ in analogy with $\omega_M$. We will also need the inverse supervielbeine, which satisfy
\be
{E_A}^{\ul{A}} {E_{\ul{A}}}^B = {\delta_A}^B.
\ee
This implies the $\Theta$ expansion
\be
\bes
{E_{\ul{M}}}^M &= {e_{\ul{M}}}^M, \cr
{E_{\ul{M}}}^{{a}} &= - \left(\ov{\Theta}\omega_{\ul{M}}\right)^{{a}}  + \left(\ov{\Theta} G_{\ul{M}} \right)^{{a}}, \cr
{E_{\ul{a}}}^{{M}} &= +{i\over 2}\left(\Gamma^{{M}}\Theta\right)_{\ul{a}}, \cr
{E_{\ul{a}}}^{{a}} &={\delta_{\ul{a}}}^{{a}} -{i\over 2}\left( \ov{\Theta} \Gamma^M \right)_{\ul{a}} \left( \ov{\Theta} \omega_M \right)^a + {i \over 3}\left( \ov{\Theta} \Gamma^M \right)_{\ul{a}} \left( \ov{\Theta} G_M \right)^a \cr
& -{i\over 24}G_{NPQR}\left( \ov{\Theta}{\mathcal{R}_{LM}}^{NPQR}\right)_{\ul{a}}\left(\ov{\Theta} \Gamma^{LM}\right)^{{a}}.
\ees
\ee

The $11$-dimensional supergravity in superspace possesses superdiffeomorphism invariance as well as invariance under the local Lorentz rotations and shifts of the super-three-form potential by exact superforms. The supersymmetries of the component-field action are a subset of all three transformations rather than any one individually. For the case of $\mathcal{N} = 1$ supergravity in $4$-dimensions, this is described in~\cite{Wess:1992cp}, which refers to this combination of transformations as supergauge transformations. At low order in $\Theta$, they were described in $11$-dimensional supergravity in~\cite{Cremmer:1980ru}.

We are interested in the preserved supergauge transformations---those that leave invariant the bulk superfields. These will correspond to the supersymmetries of the component action that are preserved by a given background. In superspace, these satisfy
\be\label{superisometry}
\bes
\delta_K {E_A}^{\ul{A}} &= K^B\partial_B {E_A}^{\ul{A}} + \partial_A K^B {E_B}^{\ul{A}} + {L^{\ul{A}}}_{\ul{B}}(K) {E_A}^{\ul{B}} = 0, \cr
\delta_K C_3 &= \mathcal{L}_K C_3 + d\Lambda_2(K) = 0,
\ees
\ee
with ${L^{\ul{A}}}_{\ul{B}}$ a local Lorentz transformation and $\Lambda_2$ a super-two-form. Using the known $\Theta$ expansion of ${E_A}^{\ul{A}}$, we can use~(\ref{superisometry}) to find a $\Theta$ expansion of $K^A$. Taking a fermionic $K^A$, i.e.\ one whose $\Theta = 0$ component is fermionic, the result up to order $\Theta^2$ is:
\be\label{superkilling}
\bes
K^M &= {i \over 2} \left( \ov{\Theta} \Gamma^M \epsilon\right), \cr
K^a &= \epsilon^a +  {i \over 8} \omega_{MNP} \left( \ov{\Theta} \Gamma^M \epsilon \right) \left( \Gamma^{NP} \Theta \right)^a \cr
& \phantom{=} + {i \over 1728} G_{NPQR} \Big( 24 \left( \ov{\Theta} \Gamma^{NP} \epsilon \right) \left( \Gamma^{QR} \Theta \right)^a - 8 \left( \ov{\Theta} \Gamma^N \epsilon \right) \left( \Gamma^{PQR} \Theta \right)^a \cr
& \phantom{ \phantom{=} + {i \over 1728} G_{NPQR} \Big\{}+ \left( \ov{\Theta} \Gamma_M \epsilon \right) \left( \Gamma^{MNPQR} \Theta \right)^a + \left( \ov{\Theta} \Gamma^{LMNPQR} \epsilon \right) \left( \Gamma_{LM} \Theta \right)^a \Big).
\ees
\ee

The spinor $\epsilon$ is required to satisfy
\be\label{fluxderiv}
\partial_M \epsilon - {1 \over 4} \omega_M{}^{\ul{MN}} \Gamma_{\ul{MN}} \epsilon - {1 \over 288} G_{NPQR} \left( \Gamma_M{}^{NPQR} + 8 \delta_M{}^N \Gamma^{PQR} \right) \epsilon = 0.
\ee

In order to leave the vielbein and three-form invariant, the superisometry (\ref{superkilling}) must be accompanied by a local Lorentz rotation and a three-form gauge transformation with parameters $L_{\ul{M}\ul{N}}$ and $\Lambda_{AB}$ respectively given by (to order $\Theta^2$)
\be
L^{\ul{M}}_{\hph{\ul{M}}\ul{N}}=-\frac{i}{2}\om_{P\hph{\ul{M}}\ul{N}}^{\hph{P}\ul{M}}\lp\ov{\Theta}\G^P\e\rp-\frac{i}{12}G^{\ul{M}}_{\hph{\ul{M}}\ul{N}PQ}\lp\ov{\Theta}\G^{PQ}\e\rp-\frac{i}{288}G_{PQRS}\lp\ov{\Theta}\G_{\hph{\ul{M}}\ul{N}}^{\ul{M}\hph{\ul{N}}PQRS}\e\rp,
\ee
\be
L^{\ul{a}}_{\hph{\ul{a}}\ul{b}}=\frac{1}{4}L_{\ul{M}\ul{N}}\lp\G^{\ul{M}\ul{N}}\rp^{\ul{a}}_{\hph{\ul{a}}\ul{b}},
\ee
and
\be
\La_{MN}=\frac{i}{2}\lp\ov{\Theta}\G_{MN}\e\rp-\frac{i}{2}C_{MNP}\lp\ov{\Theta}\G^P\e\rp,
\ee
\be
\La_{Ma}=-\frac{1}{24}\lp\G^N\Theta\rp_a\lp\ov{\Theta}\G_{MN}\e\rp-\frac{1}{24}\lp\G_{MN}\Theta\rp_a\lp\ov{\Theta}\G^N\e\rp.
\ee

\newpage

\newpage
\bibliographystyle{amsunsrt-ensp}
\bibliography{master}

\ifx\undefined\bysame
\newcommand{\bysame}{\leavevmode\hbox to3em{\hrulefill}\,}
\fi
\begin{thebibliography}{10}

\bibitem{Pestun:2009nn}
Vasily Pestun, {\em {Localization of the four-dimensional N=4 SYM to a
  two-sphere and 1/8 BPS Wilson loops}}, JHEP {\bf 1212} (2012) 067, {\tt
  arXiv:0906.0638} {\tt [hep-th]}.

\bibitem{Festuccia:2011ws}
Guido Festuccia and Nathan Seiberg, {\em {Rigid Supersymmetric Theories in
  Curved Superspace}}, JHEP {\bf 1106} (2011) 114, {\tt arXiv:1105.0689} {\tt
  [hep-th]}.

\bibitem{Becker:1996gj}
Katrin Becker and Melanie Becker, {\em {M theory on eight manifolds}}, Nucl.
  Phys. {\bf B477} (1996) 155--167, {\tt arXiv:hep-th/9605053} {\tt [hep-th]}.

\bibitem{Dasgupta:1999ss}
K.~Dasgupta, G.~Rajesh, and S.~Sethi, {\em M theory, orientifolds and
  {G}-flux}, JHEP {\bf 08} (1999) 023, {\tt arXiv:hep-th/9908088}.

\bibitem{Maldacena:1997de}
Juan~Martin Maldacena, Andrew Strominger, and Edward Witten, {\em {Black hole
  entropy in M-theory}}, JHEP {\bf 12} (1997) 002, {\tt arXiv:hep-th/9711053}.

\bibitem{Minasian:1999qn}
Ruben Minasian, Gregory~W. Moore, and Dimitrios Tsimpis, {\em Calabi-yau black
  holes and (0,4) sigma models}, Commun. Math. Phys. {\bf 209} (2000) 325--352,
  {\tt hep-th/9904217}.

\bibitem{Gadde:2013sca}
Abhijit Gadde, Sergei Gukov, and Pavel Putrov, {\em {Fivebranes and
  4-manifolds}}, {\tt arXiv:1306.4320} {\tt [hep-th]}.

\bibitem{Gadde:2013lxa}
\bysame, {\em {(0, 2) trialities}}, JHEP {\bf 03} (2014) 076, {\tt
  arXiv:1310.0818} {\tt [hep-th]}.

\bibitem{Benini:2013cda}
Francesco Benini and Nikolay Bobev, {\em {Two-dimensional SCFTs from wrapped
  branes and c-extremization}}, JHEP {\bf 06} (2013) 005, {\tt arXiv:1302.4451}
  {\tt [hep-th]}.

\bibitem{Bak:2015taa}
Dongsu Bak and Andreas Gustavsson, {\em {Partially twisted superconformal M5
  brane in R-symmetry gauge field backgrounds}}, {\tt arXiv:1508.04496} {\tt
  [hep-th]}.

\bibitem{Kallosh:2005yu}
Renata Kallosh and Dmitri Sorokin, {\em Dirac action on m5 and m2 branes with
  bulk fluxes}, JHEP {\bf 05} (2005) 005, {\tt hep-th/0501081}.

\bibitem{Martucci:2005rb}
Luca Martucci, Jan Rosseel, Dieter Van~den Bleeken, and Antoine Van~Proeyen,
  {\em {Dirac actions for D-branes on backgrounds with fluxes}}, Class. Quant.
  Grav. {\bf 22} (2005) 2745--2764, {\tt arXiv:hep-th/0504041} {\tt [hep-th]}.

\bibitem{Tsimpis:2007sx}
Dimitrios Tsimpis, {\em {Fivebrane instantons and Calabi-Yau fourfolds with
  flux}}, JHEP {\bf 0703} (2007) 099, {\tt arXiv:hep-th/0701287} {\tt
  [hep-th]}.

\bibitem{Butter:2015tra}
Daniel Butter, Gianluca Inverso, and Ivano Lodato, {\em {Rigid 4D $
  \mathcal{N}=2 $ supersymmetric backgrounds and actions}}, JHEP {\bf 09}
  (2015) 088, {\tt arXiv:1505.03500} {\tt [hep-th]}.

\bibitem{Bergshoeff:1999db}
Eric Bergshoeff, Ergin Sezgin, and Antoine Van~Proeyen, {\em {(2,0) tensor
  multiplets and conformal supergravity in D = 6}}, Class. Quant. Grav. {\bf
  16} (1999) 3193--3206, {\tt arXiv:hep-th/9904085} {\tt [hep-th]}.

\bibitem{Cordova:2013cea}
Clay Cordova and Daniel~L. Jafferis, {\em {Complex Chern-Simons from M5-branes
  on the Squashed Three-Sphere}}, {\tt arXiv:1305.2891} {\tt [hep-th]}.

\bibitem{Bah:2012dg}
Ibrahima Bah, Christopher Beem, Nikolay Bobev, and Brian Wecht, {\em
  {Four-Dimensional SCFTs from M5-Branes}}, JHEP {\bf 06} (2012) 005, {\tt
  arXiv:1203.0303} {\tt [hep-th]}.

\bibitem{Sethi:2007bw}
S.~Sethi, {\em A note on heterotic dualities via {M}-theory}, Phys. Lett. B
  {\bf 659} (2008) 385--387, {\tt arXiv:0707.0295} {\tt [hep-th]}.

\bibitem{McOrist:2010jw}
Jock McOrist, David~R. Morrison, and Savdeep Sethi, {\em {Geometries,
  Non-Geometries, and Fluxes}}, {\tt arXiv:1004.5447} {\tt [hep-th]}.

\bibitem{Andriot:2011iw}
David Andriot, {\em {Heterotic string from a higher dimensional perspective}},
  {\tt arXiv:1102.1434} {\tt [hep-th]}.

\bibitem{Gu:2014ova}
Jie Gu and Hans Jockers, {\em {Nongeometric F-theory--heterotic duality}},
  Phys. Rev. {\bf D91} (2015) 086007, {\tt arXiv:1412.5739} {\tt [hep-th]}.

\bibitem{Malmendier:2014uka}
Andreas Malmendier and David~R. Morrison, {\em {K3 surfaces, modular forms, and
  non-geometric heterotic compactifications}}, Lett. Math. Phys. {\bf 105}
  (2015) 1085--1118, {\tt arXiv:1406.4873} {\tt [hep-th]}.

\bibitem{Hull:2009mi}
Chris Hull and Barton Zwiebach, {\em {Double Field Theory}}, JHEP {\bf 09}
  (2009) 099, {\tt arXiv:0904.4664} {\tt [hep-th]}.

\bibitem{Aldazabal:2013sca}
Gerardo Aldazabal, Diego Marques, and Carmen Nunez, {\em {Double Field Theory:
  A Pedagogical Review}}, Class. Quant. Grav. {\bf 30} (2013) 163001, {\tt
  arXiv:1305.1907} {\tt [hep-th]}.

\bibitem{Gauntlett:2001ur}
Jerome~P. Gauntlett, Nakwoo Kim, Dario Martelli, and Daniel Waldram, {\em
  {Five-branes wrapped on SLAG three cycles and related geometry}}, JHEP {\bf
  11} (2001) 018, {\tt arXiv:hep-th/0110034} {\tt [hep-th]}.

\bibitem{Triendl:2015lka}
Hagen Triendl, {\em {Supersymmetric branes on curved spaces and fluxes}}, {\tt
  arXiv:1509.02926} {\tt [hep-th]}.

\bibitem{Witten:1996md}
Edward Witten, {\em {On flux quantization in M theory and the effective
  action}}, J.Geom.Phys. {\bf 22} (1997) 1--13, {\tt arXiv:hep-th/9609122} {\tt
  [hep-th]}.

\bibitem{Sethi:1996es}
S.~Sethi, C.~Vafa, and Edward Witten, {\em Constraints on low-dimensional
  string compactifications}, Nucl. Phys. {\bf B480} (1996) 213--224, {\tt
  hep-th/9606122}.

\bibitem{Berg:2002es}
Marcus Berg, Michael Haack, and Henning Samtleben, {\em {Calabi-Yau fourfolds
  with flux and supersymmetry breaking}}, JHEP {\bf 04} (2003) 046, {\tt
  arXiv:hep-th/0212255} {\tt [hep-th]}.

\bibitem{Prins:2013wza}
Dani{\"e}l Prins and Dimitrios Tsimpis, {\em {Type IIA supergravity and M
  -theory on manifolds with SU(4) structure}}, Phys.Rev. {\bf D89} (2014)
  064030, {\tt arXiv:1312.1692} {\tt [hep-th]}.

\bibitem{Gauntlett:2002sc}
Jerome~P. Gauntlett, Dario Martelli, Stathis Pakis, and Daniel Waldram, {\em {G
  structures and wrapped NS5-branes}}, Commun. Math. Phys. {\bf 247} (2004)
  421--445, {\tt arXiv:hep-th/0205050} {\tt [hep-th]}.

\bibitem{Gauntlett:2003cy}
J.~P. Gauntlett, D.~Martelli, and D.~Waldram, {\em {Superstrings with intrinsic
  torsion}}, Phys. Rev. D {\bf 69} (2004) 086002, {\tt arXiv:hep-th/0302158}.

\bibitem{Becker:2009df}
K.~Becker and S.~Sethi, {\em {Torsional Heterotic Geometries}}, Nucl. Phys. B
  {\bf 820} (2009) 1--31, {\tt arXiv:0903.3769} {\tt [hep-th]}.

\bibitem{Cherkis:1997bx}
Sergey Cherkis and John~H. Schwarz, {\em Wrapping the m theory five-brane on
  k3}, Phys. Lett. {\bf B403} (1997) 225--232, {\tt hep-th/9703062}.

\bibitem{Park:2009me}
Jaemo Park and Woojoo Sim, {\em {Supersymmetric Heterotic Action out of M5
  Brane}}, JHEP {\bf 0908} (2009) 047, {\tt arXiv:0905.2393} {\tt [hep-th]}.

\bibitem{Becker:2007zj}
K.~Becker, M.~Becker, and J.H. Schwarz, {\em {String theory and M-theory: A
  modern introduction}}.

\bibitem{Becker:2007ea}
Melanie Becker, Li-Sheng Tseng, and Shing-Tung Yau, {\em Heterotic
  kahler/non-kahler transitions}, {\tt arXiv:0706.4290 [hep-th]}.

\bibitem{Pasti:1997gx}
Paolo Pasti, Dmitri~P. Sorokin, and Mario Tonin, {\em {Covariant action for a D
  = 11 five-brane with the chiral field}}, Phys.Lett. {\bf B398} (1997) 41--46,
  {\tt arXiv:hep-th/9701037} {\tt [hep-th]}.

\bibitem{Bandos:1997ui}
Igor~A. Bandos, Kurt Lechner, Alexei Nurmagambetov, Paolo Pasti, Dmitri~P.
  Sorokin, and Mario Tonin, {\em {Covariant action for the superfive-brane of M
  theory}}, Phys. Rev. Lett. {\bf 78} (1997) 4332--4334, {\tt
  arXiv:hep-th/9701149} {\tt [hep-th]}.

\bibitem{Claus:1997cq}
Piet Claus, Renata Kallosh, and Antoine Van~Proeyen, {\em {M five-brane and
  superconformal (0,2) tensor multiplet in six-dimensions}}, Nucl. Phys. {\bf
  B518} (1998) 117--150, {\tt arXiv:hep-th/9711161} {\tt [hep-th]}.

\bibitem{Wess:1992cp}
J.~Wess and J.~Bagger, {\em {Supersymmetry and supergravity}}, 1992.

\bibitem{Becker:1995kb}
Katrin Becker, Melanie Becker, and Andrew Strominger, {\em Five-branes,
  membranes and nonperturbative string theory}, Nucl. Phys. {\bf B456} (1995)
  130--152, {\tt hep-th/9507158}.

\bibitem{Harvey:1982xk}
R.~Harvey and H.~B. Lawson, Jr., {\em {Calibrated geometries}}, Acta Math. {\bf
  148} (1982) 47.

\bibitem{Gutowski:1999tu}
J.~Gutowski, G.~Papadopoulos, and P.~K. Townsend, {\em {Supersymmetry and
  generalized calibrations}}, Phys. Rev. {\bf D60} (1999) 106006, {\tt
  arXiv:hep-th/9905156} {\tt [hep-th]}.

\bibitem{Gauntlett:2003di}
Jerome~P. Gauntlett, {\em {Branes, calibrations and supergravity}}, {Strings
  and geometry. Proceedings, Summer School, Cambridge, UK, March 24-April 20,
  2002}, 2003, pp.~79--126, {\tt arXiv:hep-th/0305074} {\tt [hep-th]}.

\bibitem{Becker:1996ay}
Katrin Becker, Melanie Becker, David~R. Morrison, Hirosi Ooguri, Yaron Oz, and
  Zheng Yin, {\em {Supersymmetric cycles in exceptional holonomy manifolds and
  Calabi-Yau 4 folds}}, Nucl. Phys. {\bf B480} (1996) 225--238, {\tt
  arXiv:hep-th/9608116} {\tt [hep-th]}.

\bibitem{Kallosh:1997ky}
Renata Kallosh, {\em {World volume supersymmetry}}, Phys.Rev. {\bf D57} (1998)
  3214--3218, {\tt arXiv:hep-th/9709069} {\tt [hep-th]}.

\bibitem{Harvey:1999as}
Jeffrey~A. Harvey and Gregory~W. Moore, {\em {Superpotentials and membrane
  instantons}}, {\tt arXiv:hep-th/9907026} {\tt [hep-th]}.

\bibitem{Tsimpis:2004gq}
Dimitrios Tsimpis, {\em {Curved 11D supergeometry}}, JHEP {\bf 0411} (2004)
  087, {\tt arXiv:hep-th/0407244} {\tt [hep-th]}.

\bibitem{Cremmer:1980ru}
E.~Cremmer and S.~Ferrara, {\em {Formulation of Eleven-Dimensional Supergravity
  in Superspace}}, Phys. Lett. {\bf B91} (1980) 61.

\end{thebibliography}

\end{document}